\documentclass[11pt,twoside]{article}
 
\usepackage{graphics}

\renewcommand{\dj}{\hbox{d\hskip-1.1ex{\raise0.640ex\hbox{--}}\skip 0.70ex}}

\begin{document}

\title{
\begin{flushright}
{\normalsize hep-ph/9901323} \\
{\normalsize IRB-TH-8/98} 
\end{flushright}
\vspace{3 cm}
{\bf Inclusive decays and lifetimes of doubly charmed baryons}} 

\author{B. Guberina\thanks{guberina@thphys.irb.hr}, 
 B. Meli\'{c}\thanks{melic@thphys.irb.hr},
 H. \v Stefan\v ci\'c\thanks{shrvoje@thphys.irb.hr}
}                    
\vspace{3 cm}
\date{
\centering
Theoretical Physics Division, Rudjer Bo\v{s}kovi\'{c} Institute, \\
   P.O.B. 1016, HR-10001 Zagreb, Croatia}


\maketitle

\begin{abstract}
The analysis of singly charmed hadrons has been extended to the
case of doubly charmed baryons, $\Xi_{cc}^{++}$, $\Xi_{cc}^{+}$
and $\Omega_{cc}^{+}$. Doubly charmed baryons are
described as a system containing a heavy $cc$-diquark 
and a light quark, similarly as in a heavy-light meson. This leads
to preasymptotic effects in semileptonic and nonleptonic
decays which are essentially proportional to the meson wave
function. Interplay between preasymptotic effects in
semileptonic and/or nonleptonic decay rates leads to very
clear predictions for semileptonic branching ratios and lifetimes
of doubly charmed baryons.
\end{abstract}
 



\section{Introduction}

Weak decays of heavy hadrons \cite{ShifBlok,Neu1,Bigi,pdg} 
present a very rich field of phenomena owing
to the complexity of confinement. Being the bound states of a heavy-quark
and light-constituents (mesons, singly charmed/bottom
baryons) or even of two heavy quarks and one light constituent (doubly
charmed/bottom baryons), heavy hadrons contain soft degrees
 of freedom which generate nonperturbative power corrections,
  such as  the destructive and/or constructive Pauli interference
 and the W-exchange/annihilation between the quark coming from the
heavy-quark decay and one of the light constituents.

Inclusive decay rates and lifetimes of charmed mesons were 
reasonably reliably calculated in the last decade. The overall picture
emerging is qualitatively satisfactory and the lifetime hierarchy has
been predicted for singly charmed baryons and found to be in
agreement with present experiments \cite{pdg}. Also, the difference in lifetimes (a factor of 2-3)
between $D^{+}$ and $D^{0}$ mesons, which is due to the negative Pauli interference
preasymptotic effect, was explained a long time
ago \cite{GubNuss,BilGubTra,ShifVol1}. 

The numerical calculations performed in the middle of the eighties
\cite{GubRuTra,ShifVol2} 
provided us with the predictions of a lifetime pattern which has
recently been confirmed by experiment. This success is rather
surprising since with the advent of the systematic Operator Product
Expansion (OPE) \cite{ShifBlok} and Heavy Quark Effective Theory 
(HQET) \cite{Neu2} it has become
clear that the charmed quark mass is not heavy enough for the
$m_{c}^{-1}$ expansion to be trustworthy. Nevertheless, it seems that
employing systematically field-theory methods to the very end of the
calculation (up to the hadronic wave function for which we have to
rely upon some phenomenological models), one is able to make clear
predictions to be compared with present and future experiments and
possibly to disentangle between various preasymptotic effects.

On the other hand, the inverse bottom quark mass
appears to be a good expansion parameter in bottom decays. However, the role of four
quark operators is negligible there (effects of the $O(1\%)$)
leaving charmed hadron decays as a playground for studying such
effects 
and as a test of the possible violation of the 
quark-hadron duality.

In this paper we extend the analysis of singly charmed baryons
decays and lifetimes \cite{GM} to the case of doubly charmed baryons.
Recently \cite{Kis}, a rather phenomenological approach using effective
constituent quark masses and fit of singly charmed baryon decays
has been employed to study doubly charmed baryon decays. We ,
however, have used a systematic field-theory approach to the very end
in order to be consistent with the previous treatment of singly
charmed hadrons and have also included the preasymptotic effects in
semileptonic decay rates of doubly charmed baryons, thus calculating 
all decay rates at the Cabibbo subleading level. We show that preasymptotic
effects dramatically change the simple spectator picture and lead
to a very clear pattern of semileptonic branching ratios and lifetimes.

\section{Preasymptotic effects and wave function in doubly charmed 
baryon decays}

Using the optical theorem, the inclusive decay width of a hadron
$H_{cc}$ with mass $M_{H_{cc}}$ containing two heavy $c$ quarks
can be written as

\begin{equation}
\label{eq:opt}
\Gamma(H_{cc} \rightarrow f)  = \frac{1}{2 M_{H_{cc}}} 2\, Im \langle
H_{cc}| \hat{T}|H_{cc}\rangle\,,
\end{equation}

where $\hat{T}$ is the transition operator

\begin {equation}
\label{eq:trans}
\hat{T} = i \int d^4x T\{ L_{eff}(x), L_{eff}^{\dagger}(0) \} .
\end{equation}

In the following we use the Operator Product Expansion which is based on
 the assumption that the energy release in
the decay of a $c$ quark is large enough. This implies that momenta
flowing through internal lines are also large and therefore
justify the OPE.

The general formula for the decay is given by \cite{ShifBlok,Neu1,Bigi}

\begin{eqnarray}
\label{eq:gen}
 \Gamma (H_{cc} \rightarrow f) &=& \frac{G_{\rm F}^2 m_c^5}{192 \pi^3} |V|^2 
\frac{1}{2 M_{H_{cc}}} \{ c_3^f \langle H_{cc}|\overline{c}
c|H_{cc}\rangle \\ \nonumber 
&+& c_5^f \frac{ \langle H_{cc}|\overline{c} g_s \sigma^{\mu \nu} G_{\mu \nu} c|
H_{cc}\rangle }{m_c^2} \\ \nonumber
 &+& \sum_i c_6^f \frac{ \langle H_{cc}| (\overline{c} 
\Gamma_i q)(\overline{q} \Gamma_i c)|H_{cc}\rangle }{m_c^3} + O(1/m_c^4) \}\,. 
\end{eqnarray}

Here $c_3^f$ and $c_5^f$ are Wilson coefficient functions which are known
at one-loop order and tree level, respectively \cite{ShifBlok,Neu1,Bigi}.
V represents appropriate matrix elements of the CKM matrix.

Let us calculate the semileptonic decay rates first. The main
contribution is expected to come from the quark decay-type
diagram, which is proportional to $\langle H_{cc}|\overline{c}
c|H_{cc}\rangle$ and is given, to $O(m_{c}^{2})$, as

\begin{eqnarray}
\label{eq:sldec}
\Gamma_{SL}^{dec}(H_{cc}) & = & 2 \frac{G_F^2}{192\pi^3} m_c^5 (c^2 \eta_{SL}(x) P_0(x) +
s^2 \eta_{SL}(0))  \nonumber \\
 &\times & (1 - \frac{1}{2}\frac{\mu_{\pi}^2(H_{cc})}{m_c^2}+
     \frac{1}{2}\frac{\mu_G^2(H_{cc})}{m_c^2}) \,.
\end{eqnarray}

Here $\mu_{\pi}^2(H_{cc})$ and $\mu_G^2(H_{cc})$ parametrize the
matrix elements of the kinetic energy and the chromomagnetic
operators, respectively. Their determination will be discussed
later. 

Throughout the paper we use the abbreviations $s^2$ and $c^2$ for 
$sin^2\theta_c$ and $cos^2\theta_c$ ($\theta_c$ is the Cabibbo angle).

The next contribution is coming from the dimension-five operator

\begin{equation}
\label{eq:slg}
\Gamma_{SL}^{G}(H_{cc})  = 2 \frac{G_F^2}{192\pi^3} m_c^5 (c^2  P_1(x) +
 s^2 ) (-2 \frac{\mu_G^2(H_{cc})}{m_c^2}) \,.
\end{equation}

%

Note that in both (\ref{eq:sldec}) and (\ref{eq:slg}) there is an additional factor 2
coming from the decays of {\em two} $c$ quarks in the doubly charmed
baryon.

The phase-space corrections $P_0$ and $P_1$ 
are cited explicitly in the Appendix. The radiative QCD correction 
$\eta_{SL}$ \cite{Kim,Nir} is given by

\begin{equation}
\label{eq:etasl}
\eta_{SL}(x)=1-\frac{2}{3} \frac{\alpha_{S}}{\pi} g(x) \, ,
\end{equation}

where for $g(x)$ we have

\begin{equation}
\label{eq:slmass}
g(x)=\pi^{2}-\frac{25}{4}+x(18+8\pi^{2}+24\ln x) \, ,
\end{equation}

and $x = m_s^2/m_c^2$. Leptons are taken to be massless.

Recently, Voloshin has noticed \cite{Vol} that preasymptotic effects in
{\em semileptonic} inclusive decays can be very large owing to the
constructive Pauli interference, the result up to the CKM
matrix element being given by

\begin{equation}
\label{eq:Voloshin}
\tilde{\Gamma}_{SL} = \frac{G_F^2}{12\pi} m_c^2 
(4 \sqrt{\kappa}-1) 5 |\psi(0)|^2\,. 
\end{equation}

Here $\kappa$ is a correction due to the hybrid renormalization
of the effective Lagrangian and it takes care of the evolution of 
$L_{eff}$ from $m_{c}$ down to the typical hadronic scale 
$\mu \sim 0.5-1 \,  GeV$. The factor 5 in front of $|\psi(0)|^2$
reflects the spin structure of doubly charmed baryons.
The baryon wave function $\psi(0)$ will be discussed later.


The total semileptonic rate for one lepton species is given by

\begin{equation}
\label{eq:sltot}
\Gamma_{SL}(H_{cc}) = \Gamma_{SL}^{dec}(H_{cc}) + 
\Gamma_{SL}^{G}(H_{cc}) + \Gamma_{SL}^{Voloshin}(H_{cc})\,,
\end{equation}

where 

\begin{eqnarray}
\label{eq:slVol}
\Gamma_{SL}^{Voloshin}(\Xi_{cc}^{++}) &=& 0 \,,\nonumber\\
\Gamma_{SL}^{Voloshin}(\Xi_{cc}^{+}) &=&   s^2 \tilde{\Gamma}_{SL} \,,\nonumber\\
\Gamma_{SL}^{Voloshin}(\Omega_{cc}^{+}) &=&  c^2 \tilde{\Gamma}_{SL}\,. 
\nonumber\\
\end{eqnarray}

In view of the significant preasymptotic effects in the SL decay rates
of singly charmed baryons, one can expect a large Pauli-interference 
contribution in the semileptonic decay rate of the $\Omega_{cc}^{+}$ baryon ($ccs$
quark structure), where that
contribution is present at the leading Cabibbo level.

Nonleptonic decay rates are slightly more complicated, since in the final
state the
lepton pair is substituted by a quark pair. 
The contributions analogous to (\ref{eq:sldec}) and (\ref{eq:slg}) are ( including
$O(m_{c}^{3})$ corrections)

\begin{eqnarray}
\label{eq:nldec}
\Gamma_{NL}^{dec}(H_{cc}) &=& 2 \frac{G_F^2}{192\pi^3}m_c^5
(c_{-}^2 + 2 c_{+}^2) [((c^4+s^4)P_{0}(x)+ \nonumber \\
& &c^2 s^2)\eta_{NL}(x)+
 c^2 s^2 \tilde{P}_{0}(x)\tilde{\eta}_{NL}(x)] \nonumber \\
&\times & [1 -\frac{1}{2}\frac{\mu_{\pi}^2(H_{cc})}{m_c^2} +
\frac{1}{2}\frac{\mu_{G}^2(H_{cc})}{m_c^2}]  \, ,
\end{eqnarray}
\begin{eqnarray}
\label{eq:nlg}
\Gamma_{NL}^{G}(H_{cc}) &=& 2 \frac{G_F^2}{192\pi^3}m_c^5
\{(2 c_{+}^2 +  c_{-}^2) [((c^4+s^4)P_{1}(x)+ \nonumber \\
& & c^2 s^2)\eta_{NL}(x)+
c^2 s^2 \tilde{P}_{1}(x){\tilde{\eta}}_{NL}(x)] \nonumber \\
&+& 2( c_{+}^2 - c_{-}^2)[((c^4+s^4)P_{2}(x)+ c^2 s^2)\eta_{NL}(x)+\nonumber \\
& &c^2 s^2 \tilde{P}_{2}(x)\tilde{\eta}_{NL}(x))]\}
(-2 \frac{\mu_{G}^2(H_{cc})}{m_{c}^{2}}) \, .
\end{eqnarray}

Radiative corrections to the nonleptonic decay, $\eta_{NL}(x)$ and  $\tilde{\eta}_{NL}(x)$, 
are far more complicated 
than analogous corrections (\ref{eq:etasl}) and
(\ref{eq:slmass}) to the semileptonic decay and the reader is referred 
to the original paper where they were first calculated \cite{Bagan}.

Again, the preasymptotic effects are expected to contribute 
significantly to the total nonleptonic decay rate. They are given
by

\begin{eqnarray}
\label{eq:four}
\Gamma^{ex} &=& \frac{G_F^2}{2\pi} m_c^2 [ c_{-}^2 + \frac{2}{3}(1 - 
\sqrt{\kappa})(c_{+}^2 - c_{-}^2)]\,5|\psi(0)|^2\,, \nonumber \\
\Gamma^{int}_{-} &=& \frac{G_F^2}{2\pi} m_c^2 [ -\frac{1}{2} c_{+}(2 c_{-}-
c_{+}) \nonumber \\ 
& & -  \frac{1}{6}(1 - \sqrt{\kappa})(5 c_{+}^2 + c_{-}^2-6 c_{+} c_{-})]
\,5|\psi(0)|^2 \,,\nonumber \\
\Gamma^{int}_{+} &=& \frac{G_F^2}{2\pi} m_c^2 [ \frac{1}{2} c_{+}(2 c_{-}+
c_{+})  \nonumber \\
& & -  \frac{1}{6}(1 - \sqrt{\kappa})(5 c_{+}^2 + c_{-}^2+6 c_{+} c_{-})]
\,5|\psi(0)|^2 \, .
\end{eqnarray}

An explicit calculation leads to the following nonleptonic decay
rates:

\begin{eqnarray}
\label{eq:nltot}
\Gamma_{NL}(\Xi_{cc}^{++}) & = & \Gamma_{NL}^{dec}(\Xi_{cc}^{++})+
\Gamma_{NL}^{G}(\Xi_{cc}^{++}) \nonumber \\
&+&\{(c^4+s^4)P_{int}(x) 
 +c^2s^2(1+\tilde{P}_{int}(x))\}\Gamma_{-}^{int} \, , \nonumber \\
\Gamma_{NL}(\Xi_{cc}^{+}) & = & \Gamma_{NL}^{dec}(\Xi_{cc}^{+})+
\Gamma_{NL}^{G}(\Xi_{cc}^{+}) \nonumber \\
& &+(c^4 P_{ex}(x)+c^2 s^2) \Gamma^{ex} \nonumber \\
& & +(s^4 P_{int}(x)+c^2 s^2)\Gamma_{+}^{int} \, ,\nonumber \\
\Gamma_{NL}(\Omega_{cc}^{+}) & = & \Gamma_{NL}^{dec}(\Omega_{cc}^{+})+
\Gamma_{NL}^{G}(\Omega_{cc}^{+}) \nonumber \\
& & +(c^4+c^2 s^2 P_{int}(x))\Gamma_{+}^{int} \nonumber \\
& & +(c^2 s^2 P_{ex}(x)+s^4)  \Gamma^{ex} \, .
\end{eqnarray}

All corrections $P$ and $\tilde{P}$ are given explicitly in the Appendix.

An important remark to be made here concerns the mass parameters in
the calculation of the matrix elements $\mu_{\pi}^{2}$ and 
$\mu_{G}^{2}$. Whenever we perform an expansion which is
essentially a field-theoretic procedure (either the OPE for the transition
operator $\hat{T}$ or the HQET expansion in the case of the $\overline{c}
c$ operator), the expansion parameter is always the current heavy-quark
running mass 
$m_{c}$. On the other hand, in the calculation of the
matrix elements, which is performed within quark models, it is
more appropriate to use constituent quark masses $m^{*}$.

Following this procedure, we give the expressions for $\mu_{\pi}^{2}$
and $\mu_{G}^{2}$. For $\mu_{\pi}^{2}$, we have 

\begin{equation}
\label{eq:mupi}
\mu_{\pi}^{2}=m_{c}^{2}v_{c}^{2}=(\frac{m_{q}^{*} T}
{2 m_{c}^{*2}+m_{c}^{*} m_{q}^{*}}+\frac{T}{2 m_{c}^{*}})m_{c}^{2} \, ,
\end{equation}

where $v_{c}$ is the average heavy-quark velocity in the $ccq$ baryon,
$m_{c}^{*}$ and $m_{q}^{*}$ are constituent masses of the heavy and
the light quark, respectively, and T is the average kinetic energy of the
light quark and the heavy diquark. The precise description of this
calculation is given in \cite{Kis} and relies upon some
phenomenological features of the meson potential.

The contributions to the $\mu_{G}^{2}$ operator, connected to the
matrix element of the chromomagnetic operator, can be divided into two
parts. The first part includes effects coming from the heavy-light
chromomagnetic interaction and these contributions can also be 
found in the singly charmed baryon $\Omega_{c}^{+}$.
The second part comprises effects originating within the heavy
diquark, i.e. heavy-heavy chromomagnetic interactions. These
effects are new \cite{Kis,BenBuch} and characteristic of doubly charmed
baryons. Their estimation relies upon the nonrelativistic QCD model 
calculation \cite{Kis,BenBuch,Lep1,Lep2}. The final expression is 

\begin{equation}
\label{eq:mug}
\mu_{G}^{2}=\frac{2}{3}(M_{ccq}^{*}-M_{ccq})m_{c}-
(\frac{2}{9} g_{S}^{2} \frac{|\phi(0)|^2}{m_{c}^{*}}+
\frac{1}{3} g_{S}^{2} \frac{|\phi(0)|^2}{m_{c}}) \, ,
\end{equation}

where the first term describes the heavy diquark-light quark
hyperfine interaction, while the second and the third correspond to the 
interaction of two heavy $c$-quarks in a diquark state. They are 
of the "chromomagnetic" and "Darwin" type, respectively.  
In (\ref{eq:mug}), $M_{ccq}$ is the mass of the doubly charmed baryon, $M_{ccq}^{*}$
is the mass of its $3/2$ spin counterpart and $\phi(0)$ is the 
wavefunction of the $cc$ pair in the heavy diquark, i.e. $|\phi(0)|^2$ is the probability 
for these two heavy quarks to meet at one point.

In the calculations above, up to the hadronic matrix elements 
we have used field theory only. The results are
expressed in terms of the baryon wave function $\psi(0)
$ and the matrix elements of the kinetic and chromomagnetic
operators, which are $\mu_{\pi}^{2}$ and $\mu_{G}^{2}$, respectively. The use of the usual
singly charmed baryon wave function $\Psi(0)$, as given
in \cite{GM}, would be premature, since intuitively, one expects a 
two-heavy-quark system to behave differently from the 
single-heavy-quark one.

In the case of singly charmed baryons, the heavy quark is "sitting"
in the centre of the baryon and the other two light quarks are moving 
around. Their spin and the color charges are correlated (in order to have 
the appropriate spin and color structure of the entire baryon), but their 
spatial motion is not. 
In this way, one has a three-body picture of the baryon containing a 
single heavy quark and one should use the baryonic wave function 
$\Psi(0)$ accordingly. 

In the case of doubly charmed baryons, one assumes that two heavy quarks are 
strongly bound into a color antitriplet state. As far as the light quark 
is concerned, the bound state of two $cc$-quarks appears as a pointlike 
diquark object \cite{Wise,White,Sav}. Thus, in the heavy-quark limit, which 
can (presumably) still be 
applied in our case ($m_c > \Lambda_{QCD}$), a doubly charmed 
baryon appears to consist of a heavy diquark and a light quark forming 
a "meson" state. Therefore, one expects that the wave function of the 
doubly charmed baryon (which has to be considered as the light-quark 
wave function at the origin of the $cc$-diquark) behaves essentially as 
the mesonic wave function.

We use the derivation of hyperfine splittings of mesons calculated in 
the constituent nonrelativistic quark model by De R\'{u}jula et al \cite{Ruj,Cor} 
to obtain the following relation between the wave functions of the 
doubly charmed baryon and the D-meson:
\begin{equation}
\label{eq:psi}
|\psi(0)|^2 = \frac{2}{3} |\psi(0)|^2_D = 
\frac{2}{3} \frac{f_D^2 M_D \kappa^{-4/9}}{12}.
\end{equation}

\begin{figure}
\centerline{\resizebox{0.55\textwidth}{!}{\includegraphics{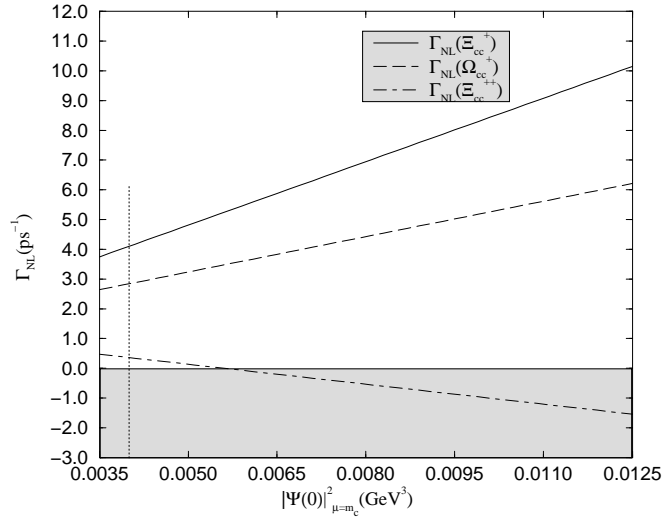}}}
\caption{Dependence of nonleptonic decay widths on the wave function squared.
The shaded area represents the unphysical region of negative $\Gamma_{NL}$ to stress
the problem of choice of the wave function. The vertical line represents the value
of the wave function squared (corresponding to $f_{D}=170 \, MeV$) 
used in the calculations.}
\end{figure}

The factor $2/3$ comes from the different spin content of doubly charmed baryons, i.e. 
the $cc$-diquark forms the spin-1 color antitriplet state.
The baryonic wave function squared in (\ref{eq:psi}) is directly proportional to the 
D-meson decay constant, $f_D$,  
squared. The factor $\kappa^{-4/9}$ is the effect of the hybrid renormalization 
which accounts for the fact that $f_D$ is measured at the scale $\sim m_c \, (\kappa =1)$, 
and one has to evolve $f_D$ down to the hadronic scale $\mu = 0.5 - 1 \, GeV$.

The choice of the mesonic wave function $\sim f_{D}$, instead of the
singly charmed baryon wave function $|\Psi(0)|^{2} \sim F_{D}^{2}$,
where $F_{D}$ is the static value of the D-meson decay constant, also seems to be consistent 
numerically. In Figure 1. we have displayed the dependence of the 
$\Gamma_{NL}(H_{cc})$ on $|\psi(0)|^{2}$ in the large 
range of the $f_D$ values. In our numerical calculation we use $f_{D}=170 \, MeV$ as a 
central value. This value is consistent both with QCD lattice 
calculations \cite{Wittig} and QCD sum rules calculations \cite{Sum1,Sum2}. In the case of
$\Gamma_{NL}(\Xi_{cc}^{++})$ there is a negative Pauli interference which
cancels the contribution coming from the decay-type diagram (\ref{eq:nldec}) and
the chromomagnetic operator (\ref{eq:nlg}). This case is very similar to the role of
the negative Pauli interference in the $D^{+}$ decay where it competes with
the decay-diagram contribution.
For $f_{D}$ large enough, the 
nonleptonic and total rates in both $D^{+}$ and $\Xi_{cc}^{++}$ 
decays become negative, Fig.1. However, a reasonable choice 
of $f_{D}$ gives positive results.

\begin{figure*}[!t]
\centerline{\resizebox{1.0\textwidth}{!}{\includegraphics{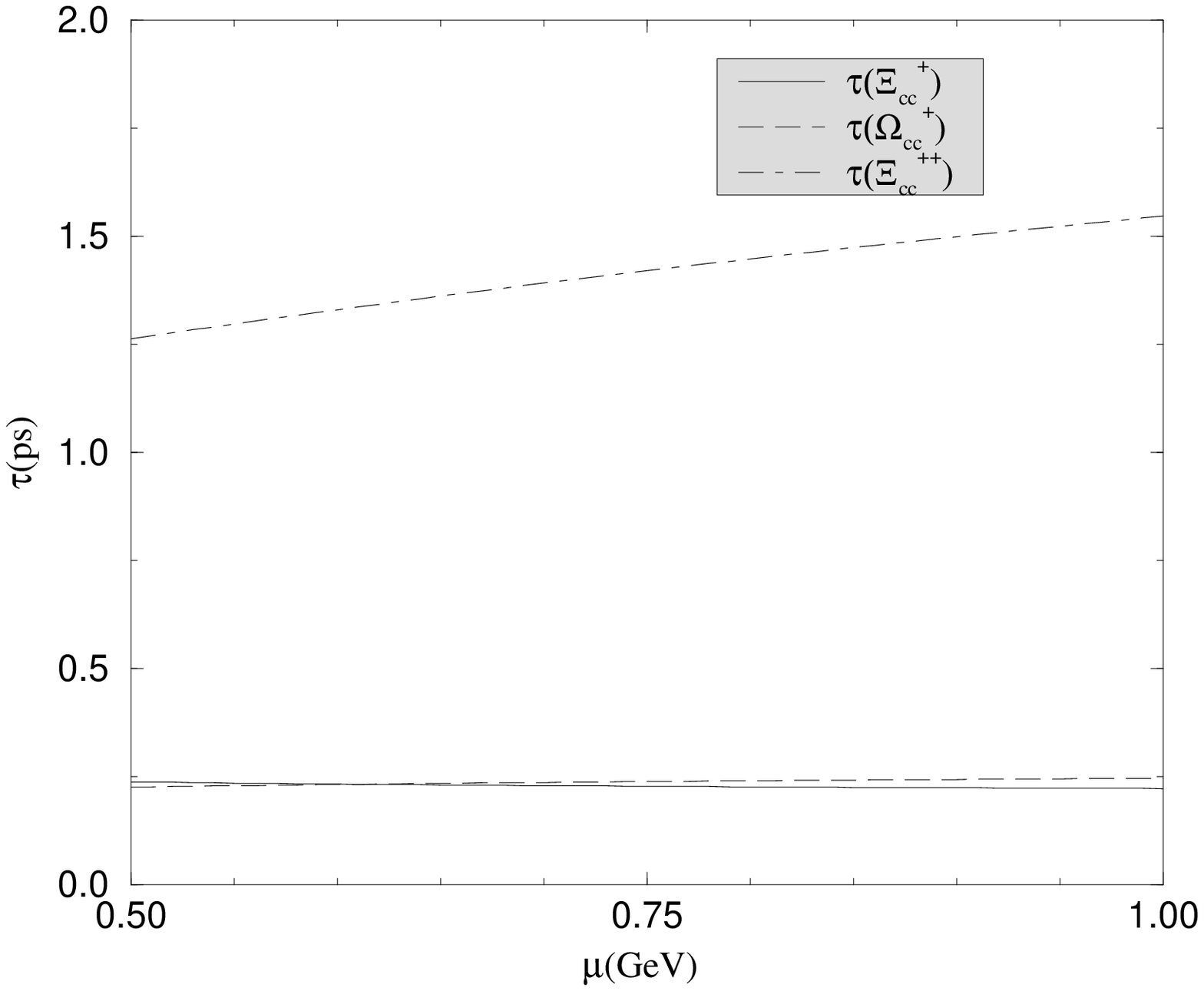} 
\includegraphics{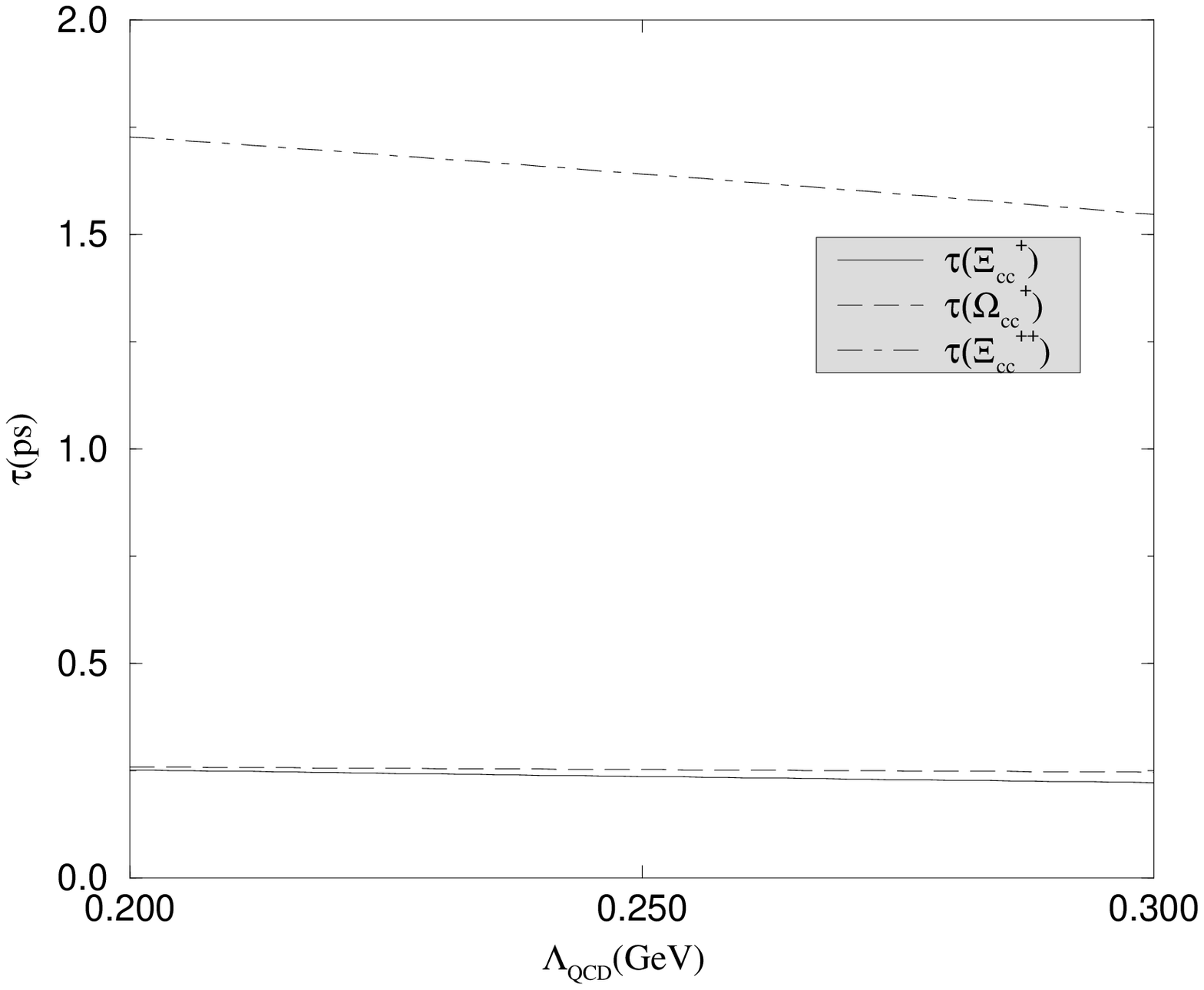}}}
\caption{Dependence of lifetimes on the parameters $\mu$ and $\Lambda_{QCD}$.
Both pictures show the evident insensitivity of $\tau(\Omega_{cc}^{+})$
and $\tau(\Xi_{cc}^{+})$ under variation of $\mu$ and $\Lambda_{QCD}$, and a 
small but notable sensitivity of $\tau(\Xi_{cc}^{++})$.}
\end{figure*}

In view of these facts and results, we may conclude that 
the phenomenological rule of using $f_D$ in mesonic and its 
static value $F_{D}$ in baryonic systems, employed first in singly charmed 
hadrons \cite{ShifBlok}, can be successfully extended to the consideration of 
doubly charmed baryons. Not doing so, but taking $F_{D}$ instead of $f_{D}$ would
lead us to the unphysical region.

So, our result is a confirmation of the above mentioned phenomenological rule
at the same, qualitative level. Taking this rule as
postulated for singly charmed hadrons, we can interpret our results 
as an extension of the same rule into the doubly charmed sector. Also,  
we can generalize the rule to some extent. We see that the use of $f_{D}$
is required in singly charmed mesons and doubly charmed baryons,
while $F_{D}$ is used in singly charmed baryons \cite{GM}. So, it is allowed 
to say that $f_{D}$ should be used in systems with two-body dynamics
(in the $ccq$ baryon case, a heavy diquark and a light quark) and $F_{D}$ in 
systems with three-body dynamics. It is important to stress that 
these considerations and conclusions are of purely phenomenological
origin, i.e. they have no direct justification in field theory.

\section{Semileptonic inclusive rates and lifetimes - results
and discussions}

In numerical calculations we use the following set of parameters, 
which closely follows the set used in \cite{GM}. For $\Lambda_{QCD}=300 \, MeV$,
the Wilson coefficients are $c_{+}=0.73$ and $c_{-}=1.88$. 
The charmed quark mass is taken to be $m_{c}=1.35 \, GeV$ and 
for the strange quark mass we use $m_{s}=150 \, MeV$.
The value of the average kinetic energy $T$, appearing explicitly in
(\ref{eq:mupi}),
is taken from \cite{Kis} to be $T = 0.4 \, GeV$, and the 
light and heavy-quark constituent masses are 
$m^{*}_q = 0.3 \, GeV$ and $m_{c}^{*}=1.6 \, GeV$, respectively.
The numerical value of the diquark wave function is also taken from
\cite{Kis} to be $|\phi(0)|=0.17 \, GeV^{3/2}$. The numerical values
for masses of doubly charmed baryons are taken from \cite{Ebert}.

As far as the $\Lambda_{QCD}$ dependence is concerned, in the range
$\Lambda_{QCD} \sim 200-300 \, MeV$ the lifetimes of $\Xi_{cc}^{+}$ and
$\Omega_{cc}^{+}$ are practically constant and the lifetime of 
$\Xi_{cc}^{++}$ is more sensitive to the value of $\Lambda_{QCD}$,
being somewhat $(10\%)$ larger for $\Lambda_{QCD}=200 \, MeV$. The same 
is true for the $\mu$-dependence in the reasonable range
$\mu \sim 0.5-1 \, GeV$. The lifetimes of $\Xi_{cc}^{+}$ and $\Omega_{cc}^{+}$
stay almost constant with variation of $\mu$, and the lifetime of
$\Xi_{cc}^{++}$ grows slowly with $\mu$ ( by $18\%$), Fig.2.

\begin{table}
\centering
\begin{tabular}{|c|c|c|c|} \hline \hline
& $  \;\;\;\;\;\;\;\;\;\Large{ \Xi_{cc}^{++}} \;\;\;\;\;\;\;\;\;$
& $\;\;\;\;\;\;\;\;\; \Large{\Xi_{cc}^{+}} \;\;\;\;\;\;\;\;\;$
 & $\;\;\;\;\;\;\;\Large{\Omega_{cc}^{+}} \;\;\;\;\;\;\;$ \\ \hline \hline
\multicolumn{4}{|c|}{Nonleptonic widths in $ps^{-1}$ } \\ \hline
$ \Gamma_{NL}$ & 0.345 & 4.158 & 2.859 \\ \hline \hline
\multicolumn{4}{|c|}{Semileptonic widths in  $ps^{-1}$ }  \\ \hline
$\Gamma_{SL}$ & 0.151 &   0.173
  &  0.603 \\ \hline \hline
\multicolumn{4}{|c|}{Semileptonic branching ratios in \%} \\ \hline
$BR_{SL}$ &  23.4    & 3.9
 & 14.9 \\  \hline \hline
\multicolumn{4}{|c|}{Lifetimes in ps} \\ \hline
$\tau$ &   1.55  &  0.22
&  0.25 \\ \hline \hline
\end{tabular}
\caption{Predictions for nonleptonic widths, semileptonic widths, semileptonic branching
ratios (for one lepton species) and lifetimes of doubly charmed baryons for the values of
parameters $m_{c}=1.35 GeV$, $\mu=1 GeV$, $\Lambda_{QCD}=300 MeV$, $f_{D}=170
MeV$.}
\end{table}

\begin{figure*}[!t]
\centerline{\resizebox{1.0\textwidth}{!}{\includegraphics{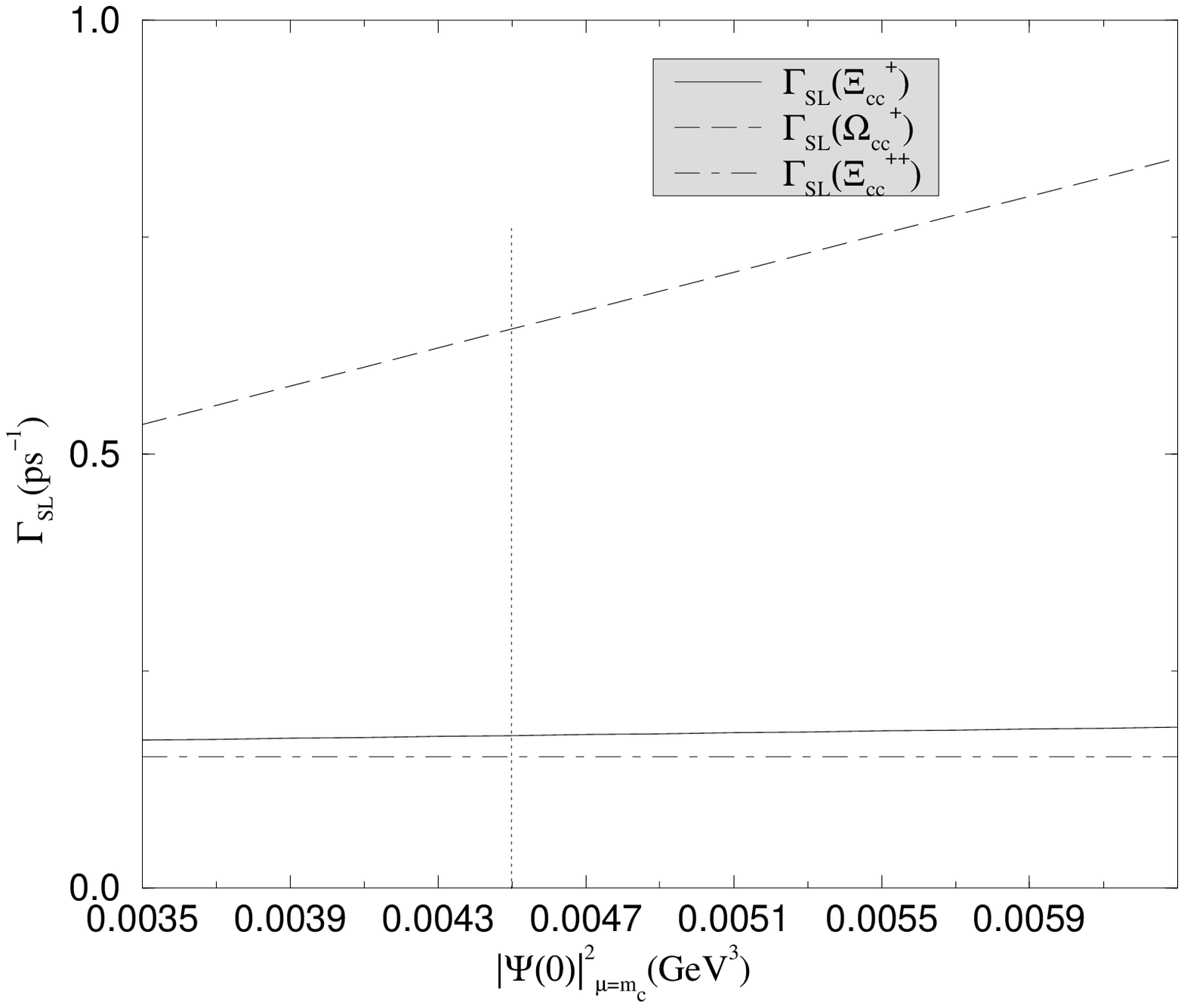} \includegraphics{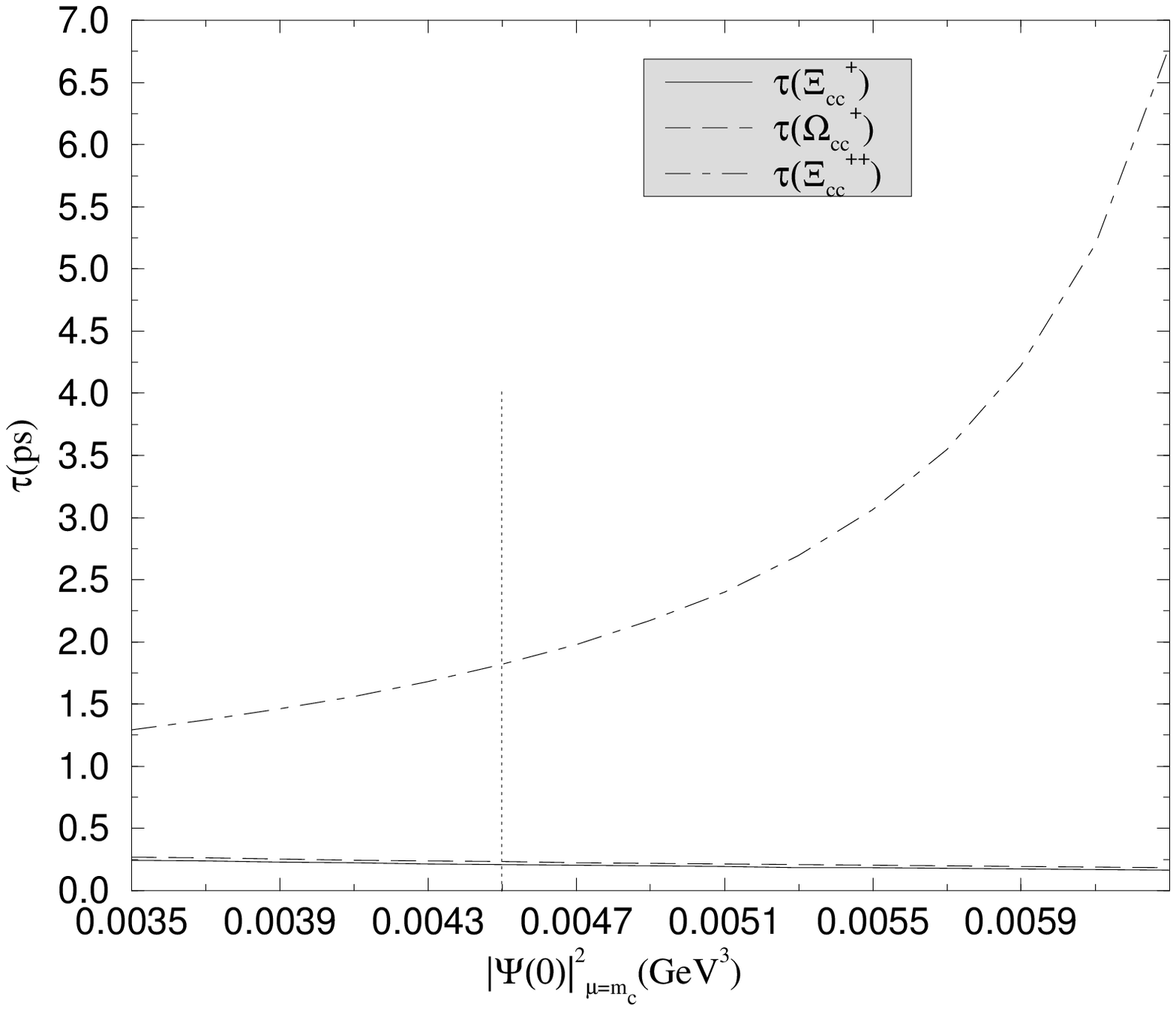}}}
\caption{Dependence of semileptonic decay widths and lifetimes on the value of the wave function squared. The vertical line represents the $\mid \psi(0) \mid^2$
used in calculations which corresponds to $f_{D}=170 MeV$.
The second picture shows the instability of 
$\tau(\Xi_{cc}^{++})$ for large $|\psi(0)|^2$.}
\end{figure*}

From Table 1 one can see that the Voloshin type of 
preasymptotic corrections in the semileptonic decay rates of $\Omega_{cc}^{+}$ 
is significant, contributing at the Cabibbo leading level, Eq.~(\ref{eq:slVol}). This contribution brings $\Gamma_{SL}(\Omega_{cc}^{+})$ to 
be four times larger than $\Gamma_{SL}(\Xi_{cc}^{++})$, which receives 
contributions only from (\ref{eq:sldec}) and (\ref{eq:slg}) . In the $\Gamma_{SL}(\Xi_{cc}^{+})$ there 
is the Pauli interference effect at the Cabibbo suppressed level, but it still 
makes the rate larger by $15\%$ than that for the $\Xi_{cc}^{++}$ baryon.

Clearly, since both semileptonic and nonleptonic rates are significantly affected by the
large preasymptotic effects which are proportional to $|\psi(0)
|^2 \sim f_{D}^{2}$, the results for lifetimes and the semileptonic branching
ratio for $\Omega_{cc}^{+}$ depend crucially on the choice of $f_{D}$. 
The latter is obvious from Fig.3, where $\Gamma_{SL}(\Omega_{cc}^{+})$
grows linearly with $f_{D}^{2}$ and especially from the second
picture in Fig. 3, where it is clear that $\tau(\Xi_{cc}^{++})$
shows instability for $f_{D}$ larger than $180 \, MeV$ owing to the large
cancelation between the negative Pauli interference term and the contributions from Eqs. 
(\ref{eq:nldec}) and (\ref{eq:nlg}). This is a clear signal that one should not take the 
results for $\Xi_{cc}^{++}$ too literally.

Keeping the above remarks in mind, we predict the following
pattern for semileptonic branching ratios:

\begin{equation}
\label{eq:brsl}
BR_{SL}(\Xi_{cc}^{+}) \ll BR_{SL}(\Omega_{cc}^{+}) \ll
BR_{SL}(\Xi_{cc}^{++}) \, , 
\end{equation}

and the following pattern for the lifetimes:

\begin{equation}
\label{eq:tau}
\tau(\Xi_{cc}^{+}) \sim \tau(\Omega_{cc}^{+}) \ll
\tau(\Xi_{cc}^{++}) \,.
\end{equation}

We can compare our results with the calculations of lifetimes 
of $\Xi_{cc}^{++}$ and $\Xi_{cc}^{+}$, which appeared recently \cite{Kis}.
The authors of that paper employed a similar field-theory technique,
but had a different approach to the choice of relevant parameters.
Throughout their paper they used the constituent heavy-quark mass as an expansion
parameter, which is a phenomenological procedure that we do not find
fully consistent. In the calculation of semileptonic decay rates,
they did not include large preasymptotic effects which significantly
change total semileptonic widths. 
Comparison shows that our numerical results are significantly different
from those of Kiselev et. al. \cite{Kis}.

Finally, it is worth discussing briefly the decay of the 
heaviest weakly decaying charmed hadron, the triply charmed baryon $\Omega_{ccc}^{++}$.
Although its complicated structure and its intrinsic tree-body motion prevent us from 
applying to this particle the heavy-light picture as described above 
to other weakly decaying heavy hadrons, 
it is possible 
to give some qualitative predictions for the $\Omega_{ccc}^{++}$ decay rate and the 
lifetime. In this baryon, preasymptotic effects (giving large contributions in
the singly and the doubly case) do not exist for lack of light valence quarks. Thus,
the dominant contribution comes from the operators of dimensions three and five. Since in doubly
 charmed decays the contribution of dimension-five operators represents less than $20\%$
of the contribution of the decay (dimension three) operator, it seems reasonable to approximate
the total decay width of $\Omega_{ccc}^{++}$ with the triple $c$-quark decay contribution
and estimate the error of disregarding dimension-five operators at the level of $20\%$.
In this case, the expression for $\Gamma_{TOT}(\Omega_{ccc}^{++})$ can be obtained
by multiplying the expressions (\ref{eq:sldec}), (\ref{eq:slg}), (\ref{eq:nldec}) and
(\ref{eq:nlg}) by a factor of $3/2$, summing them and taking the limit $\mu_{\pi}^{2}
\rightarrow 0$ and $\mu_{G}^{2} \rightarrow 0$. The numerical value for the lifetime is

\begin{equation}
\tau(\Omega_{ccc}^{++})=0.43 \, ps \,.
\end{equation}

As the cal\-cu\-la\-tion of dimen\-sion-five opera\-tor contri\-bu\-tions in triply charm\-ed 
baryon decay rates is out of scope of
the present paper, this result can be considered only as 
qualitatively correct.

\section{Conclusions}

Application of the heavy-quark expansion to the problem of inclusive 
decays of doubly charmed baryons enables us to give very interesting
predictions for their lifetimes and semileptonic branching ratios. 
Large lifetime differences are present between $\Xi_{cc}^{++}$ on the 
one hand and $\Xi_{cc}^{+}$ and $\Omega_{cc}^{+}$ on the other. Our
numerical results pick out $\Xi_{cc}^{++}$ as the longest living charmed
particle (Fig.4), although the numerical value for $\tau(\Xi_{cc}^{++})$
should be taken with certain reserve for reasons already
mentioned. Such a large numerical difference within the lifetime
hierarchy makes these predictions suitable for testing by the
first forthcoming experimental observation of doubly charmed baryons.
A theoretical prediction for semileptonic branching ratios is even clearer and the
hierarchy of $BR_{SL}$ is unambiguously determined.

\begin{figure}
\centerline{\resizebox{0.55\textwidth}{!}{\includegraphics{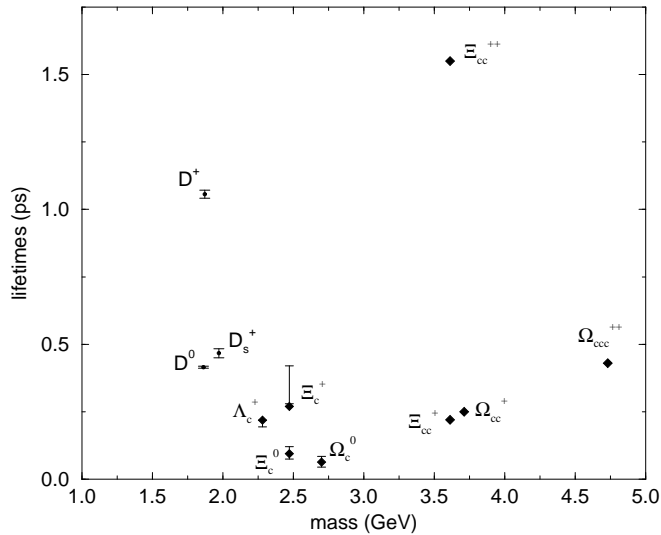}}}
\caption{Dependence of lifetimes of weakly decaying charmed hadrons on their
 mass. Experimental results are drawn as points with error bars (for singly 
charmed hadrons), while theoretical predictions are drawn as filled diamonds
(all weakly decaying charmed baryons). Values for masses of doubly and triply 
charmed baryons are taken from \cite{Koer}.}
\end{figure}

The total hierarchy of lifetimes for charmed hadrons is shown in Figure 4.
It is evident that charmed hadrons show a very complex pattern in the 
$\tau - M$ plane. One can note that doubly charmed baryon lifetimes are 
comparable with those of their singly charmed counterparts. This result is just 
opposite to the naive expectation of roughly double widths in the doubly 
charmed case owing to the decay of two, instead of one $c$ quark, or, 
correspondingly twice smaller lifetimes of $ccq$ baryons. 
However, the mesonic nature of doubly charmed baryon wave functions, where one
uses the smaller $f_D$ constant instead of its static $F_D$ value, reduces 
four-quark operator contributions, increasing doubly charmed baryon lifetimes.

Although the $c$ quark is considered not being heavy enough to ensure
a reasonable convergence of the heavy-quark expansion series, 
the numerical results in the singly charmed
sector show satisfactory qualitative and even quantitative
agreement with experiment \cite{GM}. If future experiments concerning the doubly
(and triply) charmed sector should show similar agreement with our theoretical
predictions, that would have important implications on the role of
four-quark operators and the entire theory of heavy quark expansion.
Besides, the absence or the presence of agreement might have notable implications on
the validity of the quark-hadron duality in the charmed sector.


\appendix
\begin{appendix}
\section{Appendix: Phase-space corrections}

Phase-space corrections can be understood as reduction of particle
phase space due to the propagation of the massive particle in the loops of
diagrams describing inclusive decays. In our case, we consider the $s$
quark to be massive, while the other particles ($u$,$d$ quarks and leptons)
are treated as massless. These corrections can be classified
according to the type of the the operator diagram in which they appear and
according to the number of massive quarks in the loop (or in a final
state if we consider an inclusive process as a sum of exclusive channels).

First, we shall enumerate corrections that appear in decay and 
dimension-five operator diagrams. From here 
on, $x=m_{s}^{2}/m_{c}^{2}$.

\begin{itemize}
\item One massive quark in the loop:

\begin{equation}
\label{eq:p0}
P_{0}(x)= (1-x^2)(1-8x+x^2)-12x^2\,ln\,x \,,
\end{equation}

\begin{equation}
\label{eq:p1}
P_{1}(x)=(1-x)^4\,,
\end{equation}

\begin{equation}
\label{eq:p2}
P_{2}(x)=(1-x)^3\,.
\end{equation}

$P_{0}(x)$ appears as a correction to the decay-type diagram, while $P_{1}(x)$
and $P_{2}(x)$ come as corrections to the chromomagnetic operator.

\item Two massive quarks in the loop:

Using the notation 
\begin{equation}
\label{eq:v}
v(x)=\sqrt{1-4x} \,,
\end{equation}

we have
\begin{eqnarray}
\label{eq:p00}
\tilde{P}_{0}(x) & = & v(x)(1-14x-2x^2-12x^3)+ \nonumber \\
& & 24x^2 (1-x^2) \,ln\, (\frac{1+v(x)}{1-v(x)})\, , 
\end{eqnarray}

\begin{equation}
\label{eq:p11}
\tilde{P}_{1}(x)=\frac{1}{2}(2 \tilde{P}_{0}(x)-y\partial_{y}
\tilde{P}_{0}(y) \mid_{y=x})\, ,
\end{equation}

\begin{equation}
\label{eq:p22}
\tilde{P}_{2}(x)=v(x)(3x^2+\frac{x}{2}+1)-3x(1-2x^2)\,ln\, (\frac{1+v(x)}{1-v(x)})\,.
\end{equation}

\end{itemize}

Similarly as above, 
$\tilde{P}_{0}(x)$ appears in the decay diagram, while $\tilde{P}_{1}(x)$
and $\tilde{P}_{2}(x)$ are corrections to the dimension-five operator.

Corrections in the paper due to one massive quark are systematically denoted by $P$, 
 while those due to two massive quarks are denoted by $\tilde{P}$.

Next, we display the phase-space corrections to four-quark operators:

\begin{equation}
\label{eq:pex}
P_{ex}(x)=(1-x)^2 \,,
\end{equation}

\begin{equation}
\label{eq:pint}
P_{int}(x)=(1-x)^2 (1+x) \,, 
\end{equation}

\begin{equation}
\label{eq:pheint}
\tilde{P}_{int}(x)=\sqrt{1-4x} \,.
\end{equation}

$P_{ex}(x)$ appears as a correction to the exchange diagram, $P_{int}(x)$
corrects for the massive quark in the interference contributions, while
$\tilde{P}_{int}(x)$ is a correction in the case of negative 
interference when there are two massive quarks in the loop
\cite{Bigi,Cormass,Hok}.
\end{appendix}

{\em Acknowledgements.}
This work was supported by the Ministry of Science and Technology of the Republic of
Croatia under Contract No. 00980102.


\end{document}